\documentclass[sigconf]{acmart}
\usepackage{balance}
\acmConference[LOCO ’24]{1st International Workshop on Low Carbon Computing}{December 3, 2024}{Glasgow/Online}

\acmDOI{}         % leave blank (publisher will fill)
\acmISBN{}        % leave blank (publisher will fill)

\usepackage{tabularx}
\usepackage[linesnumbered,ruled,vlined]{algorithm2e}
\usepackage{tikz}
\usetikzlibrary{positioning}

\setcopyright{rightsretained}  % Use CC BY-NC-ND license
\acmBooktitle{[LOCO ’24]{1st International Workshop on Low Carbon Computing}{December 3, 2024}{Glasgow/Online}}
\acmYear{2024}
\copyrightyear{2024}

\begin{document}

%% Use this when submitting an article to a sponsored event. You'll
%% receive a unique submission ID from the organizers
%% of the event, and this ID should be used as the parameter to this command.
%%\acmSubmissionID{123-A56-BU3}

%%
%% For managing citations, it is recommended to use bibliography
%% files in BibTeX format.
%%
%% You can then either use BibTeX with the ACM-Reference-Format style,
%% or BibLaTeX with the acmnumeric or acmauthoryear sytles, that include
%% support for advanced citation of software artefact from the
%% biblatex-software package, also separately available on CTAN.
%%
%% Look at the sample-*-biblatex.tex files for templates showcasing
%% the biblatex styles.
%%

%%
%% The majority of ACM publications use numbered citations and
%% references.  The command \citestyle{authoryear} switches to the
%% If you are preparing content for an event
%% sponsored by ACM SIGGRAPH, you must use the "author year" style of
%% the next command will enable that style.

\title{MAIZX: A Carbon-Aware Framework for Optimizing Cloud Computing Emissions}

\author{Federico Ruilova}
% \email{ruilova.alfaro@tu-berlin.de}
\email{fedra@kth.se}
\orcid{0000-0002-3577-2065}
\affiliation{%
  \institution{KTH Royal Institute of Technology}
  \city{Stockholm}
  \country{Sweden;}}

\affiliation{%
  \institution{Technische Universität Berlin}
  \city{Berlin}
  \country{Germany}
}

\author{Ernst Gunnar Gran}
\email{ernst.g.gran@ntnu.no}
\orcid{0000-0002-0349-3643}
\affiliation{%
  \institution{Norwegian University of Science and Technology (NTNU)}
  \city{Gjøvik}
  \country{Norway}}

\author{Sven-Arne Reinemo}
\email{svenar@simula.no}
\orcid{0000-0002-6167-4784}
\affiliation{%
  \institution{Simula Metropolitan Centre for Digital Engineering}
  \city{Oslo}
  \country{Norway}}

% \affiliation{%

% \affiliation{%

% \affiliation{%

% \affiliation{%

% \affiliation{%

% \affiliation{%

\renewcommand{\shortauthors}{Ruilova et al.}

\begin{abstract}
Cloud computing drives innovation but also poses significant environmental challenges due to its high energy consumption and carbon emissions. Data centers account for 2-4\% of global energy usage, and the ICT sector's share of electricity consumption is projected to reach 40\% by 2040.As the goal of achieving net-zero emissions by 2050 becomes increasingly urgent, there is a growing need for more efficient and transparent solutions, particularly for private cloud infrastructures, which are utilized by 87\% of organizations, despite the dominance of public-cloud systems.

This study evaluates the MAIZX framework, designed to optimize cloud operations and reduce carbon footprint by dynamically ranking resources, including data centers, edge computing nodes, and multi-cloud environments, based on real-time and forecasted carbon intensity, Power Usage Effectiveness (PUE), and energy consumption. Leveraging a flexible ranking algorithm, MAIZX achieved an 85.68\% reduction in CO\textsubscript{2} emissions compared to baseline hypervisor operations. Tested across geographically distributed data centers, the framework demonstrates scalability and effectiveness, directly interfacing with hypervisors to optimize workloads in private, hybrid, and multi-cloud environments. MAIZX integrates real-time data on carbon intensity, power consumption, and carbon footprint, as well as forecasted values, into cloud management, providing a robust tool for enhancing climate performance potential while maintaining operational efficiency.
\end{abstract}

% Mandatory CCS concepts for ACM papers over two pages





% \received{24 September 2024}


\maketitle

\section{Introduction}

Cloud computing drives innovation across sectors but raises concerns about its environmental impact, particularly due to high energy consumption and carbon emissions.Currently Data centers account for 2-4\% of the global energy usage, with the broader ICT sector consuming 6\%~\cite{ross_energy_2023},\cite{the_independent_global_2016},\cite{ahvar_estimating_2022}. This is projected to rise to nearly 40\% of total global electricity consumption by 2040~\cite{international_energy_agency_net_2021}, underscoring the urgent need for sustainable solutions,  carbon-aware computing and energy-aware frameworks\cite{laplante_frameworking_2023}, \cite{woodruff_when_2023},\cite{radovanovic_carbon-aware_2023}, \cite{maji_bringing_2023},\cite{lannelongue_carbon_2023}, \cite{hanafy_war_2023}. Achieving net-zero emissions by 2050 is critical for limiting global warming, and while cloud providers focus on carbon neutrality through grid carbon intensity awareness and renewable energy integration, more transparent and effective methods are needed, especially for private cloud setups, used by 87\% of organizations\cite{flexera_flexera_2024},\cite{cisco_global_2022},\cite{arora_towards_2023}.

This research explores the MAIZX framework to revisit its potential and evaluate its climate performance\cite{ruilova_alfaro_towards_2024}; the framework uses a ranking algorithm that allocates resources based on scores of computing nodes. The framework’s scalability and effectiveness were empirically tested by implementations across geographically distributed data centers and validated via simulations. By integrating these metrics into cloud management, MAIZX offers a robust tool for enhancing climate performance and assessing environmental impact in private, hybrid and multi-cloud approaches.

\section{Background and Related Work}
\label{sec:background}

Cloud computing enables scalable and flexible computing resource allocation via Infrastructure as a Service (IaaS), Platform as a Service (PaaS), and Software as a Service (SaaS)~\cite{mell_nist_2011}. These services operate over different deployment models—public, private, hybrid, and multi-cloud—each offering varying levels of control, scalability, and resource management~\cite{stanoevska-slabeva_cloud_2010}. 

Recent industry trends indicate that 72\% of organizations prefer hybrid cloud environments, while 87\% employ a multi-cloud approach, balancing performance, cost, and regulatory compliance~\cite{flexera_flexera_2024,cisco_global_2022}. However, rising energy demands associated with cloud infrastructures, particularly those running artificial intelligence (AI) workloads, have amplified concerns about sustainability~\cite{wadhwani_how_2025, bray_how_2025}. AI-driven data centers could consume up to 8\% of global electricity by 2030, with private cloud deployments accounting for a significant share due to their role in enterprise data governance~\cite{wadhwani_how_2025}.

As a result, carbon-aware computing has emerged as a crucial paradigm, integrating energy efficiency with dynamic workload allocation~\cite{arora_towards_2023}. Public cloud providers, such as AWS, Google Cloud, and Microsoft Azure, have introduced carbon tracking services, yet private and hybrid clouds lack similar transparency and adaptation mechanisms~\cite{arora_towards_2023,maji_bringing_2023}. Addressing this gap, agent-oriented frameworks offer a promising direction, enabling autonomous, AI-driven energy management~\cite{ruilova_alfaro_towards_2024}.

To mitigate cloud computing's environmental impact, carbon-aware computing seeks to align workload scheduling with energy grid conditions, optimizing operations based on real-time carbon intensity data~\cite{wiesner_cucumber_2022, radovanovic_carbon-aware_2023}. Strategies include temporal shifting of workloads to periods of lower emissions~\cite{wiesner_fedzero_2024, james_low_2019}, geographic scheduling by allocating workloads to data centers in regions with cleaner energy sources~\cite{eilam_towards_2023, radovanovic_carbon-aware_2023}, and intelligent scaling mechanisms that dynamically adjust compute resources~\cite{kim_greenscale_2023, subramanian_carbon-aware_2023}.

Building upon prior research, the MAIZX framework introduces an agent-based ranking mechanism designed for private, hybrid, and multi-cloud infrastructures~\cite{ruilova_alfaro_towards_2024}. Unlike conventional carbon-aware strategies, MAIZX operates across multiple levels of abstraction, integrating agent-driven resource allocation, multi-cloud adaptability, and real-time energy forecasting in coordination with the hypervisor of the cloud where it runs. The ranking MAIZX ranking algorithm dynamically scores computing nodes based on carbon footprint, CPU efficiency, and workload compatibility, while the agent-oriented approach enables autonomous decision-making in cloud scheduling. MAIZX also integrates with hypervisor-based schedulers, leveraging the hypervisor’s capabilities to optimize workload distribution and also interconnect with other hybrid approaches such as multicloud or hybryd (public and private)·

Preliminary studies indicate that MAIZX can reduce emissions by up to 85.68\%, outperforming baseline cloud scheduling approaches in carbon-aware workload distribution~\cite{ruilova_alfaro_towards_2024}. This aligns with broader industry trends, where AI-based cloud optimization is increasingly necessary. Training and inference of large-scale AI models, such as GPT-4, require substantial computing power, intensifying demand for sustainable AI operations~\cite{wadhwani_how_2025}. 

In order to have an idea of the impact, the concept of carbon performance potential (CPP) provides a framework for evaluating the capability of organizations and technologies to manage carbon emissions effectively while optimizing economic outputs. Historically evolving from simplistic assessments to sophisticated multi-indicator metrics, CPP has become essential in guiding organizations toward a low-carbon economy~\cite{mporg_why_2024,panwar_systematic_2022}. Recent research highlights a positive correlation between digital transformation and improved carbon management, as organizations leverage digital technologies for greater resource efficiency and sustainable innovation~\cite{eilam_towards_2023,wadhwani_how_2025}. Despite these advancements, substantial barriers remain, including technological limitations, economic constraints, regulatory uncertainty, and social resistance, underscoring the need for standardized metrics and robust policy frameworks to foster sustainable innovation and carbon reduction practices~\cite{wadhwani_how_2025}.

This research builds upon previous work in carbon-aware cloud computing by addressing key limitations in workload scheduling for private and hybrid cloud environments. The MAIZX framework leverages an agent-oriented, AI-driven approach to dynamically optimize workload distribution based on real-time carbon intensity, enhancing both sustainability and operational efficiency. Unlike existing models such as GreenScale, FedZero, and CarbonScaler—which primarily focus on different layers of abstraction~\cite{kim_greenscale_2023, wiesner_fedzero_2024, hanafy_carbonscaler_2023}—MAIZX extends these principles to private clouds, hybrid architectures, and multi-cloud federations. By integrating seamlessly with hypervisors, MAIZX’s agentic components can coordinate across diverse infrastructures, enabling adaptive workload placement and proactive energy-aware scheduling across multiple cloud environments.

\section{MAIZX Ranking Algorithm}
The MAIZX framework uses a hybrid architecture that centralizes control while distributed agents collect power consumption and carbon intensity data, both at each distributed node and at the core \cite{ruilova_alfaro_towards_2024}.

\begin{figure}[htbp]
    \centering
    \includegraphics[width=\columnwidth]{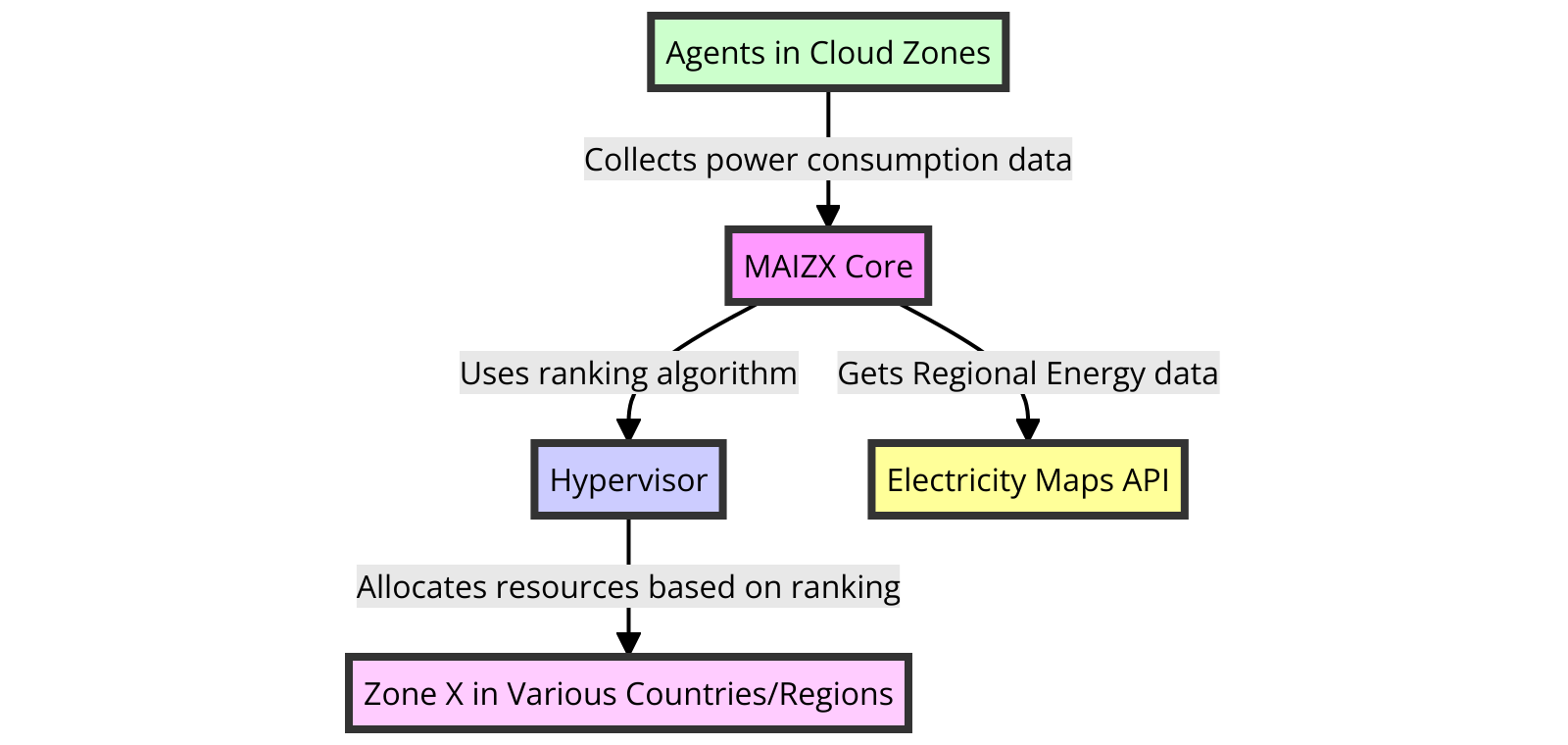}
    \Description{Diagram of the MAIZX Framework Architecture.}
    \caption{MAIZX Framework Architecture}
    \label{fig:maizx_framework}
\end{figure}

The ranking algorithm dynamically allocates workloads to nodes with the lowest carbon intensity, prioritizing environmental impact without compromising performance. Centralized components coordinate with the hypervisor using carbon efficiency and power data as it could be observed in Figure \ref{fig:maizx_framework}.

\subsection{Key Functionalities of Carbon-Aware allocation}
Agents gather real-time energy and carbon intensity data, supporting carbon footprint calculations and forecasting.

The algorithm evaluates nodes on carbon footprint and efficiency, optimizing workload distribution to reduce emissions. It integrates with hypervisors like OpenNebula \cite{opennebula_scheduler_2023} for efficient scheduling.

The ranking system calculates a node's efficiency by considering key environmental and operational parameters. The algorithm, $MAIZ\_RANKING$, is defined as:

% \begin{equation}
% \resizebox{1\columnwidth}{!}{$MAIZ\_RANKING = w_1 \cdot \text{CFP} + w_2 \cdot \text{FCFP} + w_3 \cdot \text{CP\_RATIO} + w_4 \cdot \text{SCHEDULE\_WEIGHT}$}
% \label{eq:maiz_ranking}
% \end{equation}

\begin{equation}
\label{eq:maiz_ranking}
\begin{aligned}
\text{MAIZ\_RANKING} =\;& w_{1}\,\mathrm{CFP} + w_{2}\,\mathrm{FCFP} \\[-0.3ex]
&\quad +\,w_{3}\,\mathrm{CP\_RATIO} + w_{4}\,\mathrm{SCHEDULE\_WEIGHT}\,.
\end{aligned}
\end{equation}

In this equation, CFP refers to the node’s Carbon Footprint, while FCFP denotes the Forecasted Carbon Footprint based on historical data. The Computing Power Ratio (CP\_RATIO) reflects the node’s energy efficiency, and the Scheduling Weight (SCHEDULE\_WEIGHT) accounts for workload priorities and deadlines. Adjustable weights ($w_1$, $w_2$, $w_3$, $w_4$) are assigned to each factor, enabling the framework to balance environmental impact, performance, and operational needs. This flexible approach ensures that MAIZX can maintain a balance between sustainability and efficiency across various cloud environments.

\section{Methodology and Experimental Design}

This study evaluates the MAIZX framework's climate performance in private and multi-cloud oriented environments by optimizing operations based on data center locations, hardware usage, and regional energy profiles. The assessment builds on prior MAIZX research\cite{ruilova_alfaro_towards_2024}, using 2022 carbon intensity data\cite{electricitymapscom_electricity_2024} to simulate various scenarios: the Baseline Scenario evenly distributes loads without any consideration of carbon intensity or footprint data, serving as a comparison for carbon footprint analysis.; Scenario A directs all computing power to the node with the lowest carbon intensity; Scenario B concentrates tasks on a single node while powering off others to measure energy savings; and Scenario C dynamically shifts loads based on daily carbon intensity fluctuations, highlighting MAIZX's adaptability for reducing emissions.

Data collection consists of power consumption measured every 20 seconds, while carbon intensity is recorded hourly across three regions: Spain, the Netherlands, and Germany. 
The carbon footprint for each node across the scenarios is calculated using a standard methodology \cite{eilam_towards_2021, gesi_gesi_2024}, applying the formula:

\begin{equation} CF = EC \times PUE \times CI \end{equation}

where \textbf{CF} is the carbon footprint, \textbf{EC} is energy consumption, \textbf{PUE} is Power Usage Effectiveness, and \textbf{CI} is carbon intensity
% The carbon footprint for each node through the differents cenarios is then calculated using a standard methodlogy. the following formula:

% \begin{equation} CF = EC \times PUE \times CI \end{equation}

% This equation is a standard method for calculating carbon footprint, commonly found in sustainability research and carbon accounting practices \cite{eilam_towards_2021}, \cite{gesi_gesi_2024}. 
% The components of the formula are defined as follows: \textbf{CF} represents the carbon footprint in grams of CO\textsubscript{2} equivalent (gCO\textsubscript{2}eq), indicating the total emissions produced. \textbf{EC} stands for energy consumption, measured in kilowatt-hours (kWh), reflecting the total energy usage of the computing node. \textbf{PUE} refers to the Power Usage Effectiveness, a ratio used to evaluate the energy efficiency of a data center by comparing the total facility energy to the energy consumed by the IT equipment. Finally, \textbf{CI} represents the carbon intensity (gCO\textsubscript{2}eq/kWh), which measures the carbon emissions per unit of electricity consumed, varying based on the energy source's carbon footprint.

In order to calculate the climate performance potential (CPP), the impact forecast tool was used, together with the corresponding EU taxonomy group for ICT. The calculation model uses functional unit or (FU) to calculate life cicle analys \cite{impact-forecastcom_impact_2024}, \cite{ecocostsvaluecom_life_2024},\cite{vogtlander_practical_2010}, \cite{valenzuela_how_2021}.

\section{Results and Analysis}

\begin{figure}[htbp] \centering \includegraphics[width=\columnwidth]{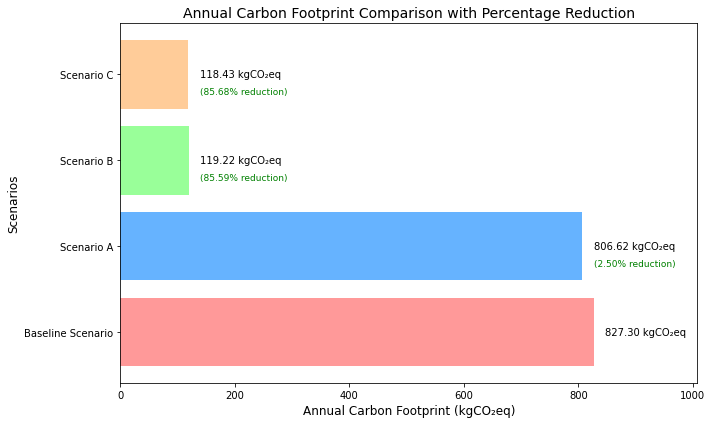} \caption{MAIZX Framework Architecture} \Description{Diagram showing the architecture of the MAIZX framework.} \label{dataloco} \end{figure}

In Scenario C (active load-shifting over one year), the MAIZX framework reduces CO\textsubscript{2} emissions by 85.68\% compared to the baseline, optimizing workloads using real-time carbon intensity data, (Figure \ref{dataloco}). Each unit, consisting of 60 servers in a 3-node private cloud, reduces emissions by 713.5 kg of CO\textsubscript{2} annually. The main difference between Scenario B and Scenario C is the use of real-time carbon data: Scenario B evenly distributes workloads without considering carbon intensity, whereas Scenario C actively shifts workloads to the nodes with the lowest carbon intensity once, and scenario A as well but leaving the other nodes available. While both scenarios B and C achieve similar reductions, Scenario C is more sustainable long-term due to its dynamic response to fluctuations in carbon intensity, consistently maintaining lower emissions when variations occur.

% To evaluate the broader impact, consider that 1\% of the EU Taxonomy targets for data-driven climate change monitoring solutions and data processing in the ICT sector, which totals 19.754 Mt CO\textsubscript{2}eq. Using this as a reference \cite{european_commission_-_european_commission_eu_2024},\cite{impact-forecastcom_impact_2024}, over a 10-year period, implementing the MAIZX framework, with its 85\% reduction capability, would result in the following reductions::

% \begin{itemize} \item Total reduction needed: 19.754 Mt CO\textsubscript{2}eq (19,754,000,000 kg). \item CO\textsubscript{2} reduction per unit: -713.5 kg per year. \item Number of units required: 27,686,054 units.(actively shifted units.)\end{itemize}

To evaluate the broader impact, if 1\% of the EU Taxonomy target for data-driven climate change monitoring and ICT data processing is considered, it totals 19.754 Mt CO\textsubscript{2}eq \cite{european_commission_-_european_commission_eu_2024},\cite{impact-forecastcom_impact_2024}. Over a 10-year period, implementing the MAIZX framework with its 85\% reduction capability would yield the following:

\begin{itemize} \item Total reduction target: 19.754 Mt CO\textsubscript{2}eq (19,754,000,000 kg). \item Annual CO\textsubscript{2} reduction per unit: 713.5 kg. \item Units required: 27,686,054. \end{itemize}
This showcases the scalability of the MAIZX framework for reducing emissions in large-scale cloud operations, potentially reaching 19.754 Mt CO\textsubscript{2}eq in 10 years targeting shifted units. The results highlight MAIZX's potential for substantial environmental and cost savings, particularly in private and multi-cloud environments with optimized power consumption and associated carbon footprint.
\section{Conclusion}

The framework demonstrates significant potential to reduce CO\textsubscript{2} emissions, particularly in private or multi-cloud setups. Despite being tested in regions with non-renewable energy matrices, MAIZX achieved considerable emission reductions. Empirical data validated the framework's carbon footprint calculations, affirming its methodology and accuracy. Conservative 10-year projections suggest that MAIZX could reduce emissions by 20 Mt CO\textsubscript{2}eq—equivalent to planting 90 million trees or removing 2.44 million cars from the road annually. Additionally, MAIZX provides significant eco-cost savings, including €3 billion in human health impacts, €4.65 billion in eco-toxicity, and €2.63 billion in carbon footprint-related costs\cite{impact-forecastcom_impact_2024}. These findings underscore MAIZX’s potential to support sustainability goals \cite{united_nations_sustainable_2023} in the ICT sector by optimizing cloud infrastructure for sustainability.

\section{Bibliography}

\bibliographystyle{ACM-Reference-Format}

\end{document}